

\vsize = 9truein 
\hsize = 6truein 
\baselineskip = 16truept 
\pageno = 1

\def\foot#1{
\footnote{($^{\the\foo}$)}{#1}\advance\foo by 1
} 


\def\IR{{\bf R}} 
\def\and{\qquad\hbox{and}\qquad}

\def\2{\hbox{$\sevenrm{1\over 2}$}}

\def\L{L}

\font\tenb=cmmib10 
\newfam\bsfam

\textfont\bsfam=\tenb

\mathchardef\betab="080C
\mathchardef\xib="0818
\mathchardef\omegab="0821
\mathchardef\deltab="080E
\mathchardef\epsilonb="080F
\mathchardef\pib="0819
\mathchardef\sigmab="081B


\newcount\ch 
\newcount\eq 
\newcount\foo 
\newcount\ref 

\def\chapter#1{
\parag\eq = 1\advance\ch by 1{\bf\the\ch.\enskip#1}
}
\def\parag{\hfil\break} 
\def\ccr{\cr\noalign{\medskip}}

\def\reference{
\parag [\number\ref]\ \advance\ref by 1
}

\ch = 0 
\foo = 1 
\ref = 1 

\centerline{NON-RELATIVISTIC CONFORMAL STRUCTURES
\foot{
Talk given at the {\rm XIX}th {\it Int. Coll. on  Group Theor. Meths. in
Physics, Salamanca'92}. Del Olmo, Santander, Guilarte (eds) p. 446,
Anales de Fisica Monografias (1993).}}
\vfill
\centerline{P. A. HORV\'ATHY}
\vskip 2mm
\centerline{D\'epartement de Math\'ematiques,
Universit\'e,}
\centerline{Parc de Grandmont,
  F-37200 TOURS (France)}

\parag
{\bf Abstract.}
{\it The ``Kaluza-Klein-type''
geometric structure appropriate to study the central extension of the
Galilei group and non-relativistic physics is reviewed.}


\chapter{Bargmann structures.}

The fundamental invariance group of
non-relativistic physics is the {\it Galilei group},
which acts on space-time
according to
$$
{\bf r}^*=A{\bf r}+{\bf b}t+{\bf c},
\eqno(1a)
$$
$$
t^*=t+e,
\eqno(1b)
$$
where $A\in SO(3)$, ${\bf b},\ {\bf c}\in\IR^3$, $e\in\IR$.
However, unlike in the relativistic case,
the Galilei group acts on the wave functions only up to a
phase [1],
$$
\psi^*({\bf r},t)=e^{-(im/\hbar)\big({\bf b.}A{\bf r}+{\bf b}^2t/2\big)}
\psi({\bf r}^*,t^*).
\eqno (2)
$$
Hence, it is only a central extension of the
  Galilei group (called the
Bargmann group), which is a symmetry group at the
quantum level.

The natural geometric setting for realizing the central
extension of a Lie group is to add a
new, \lq vertical' variable, $s$, and consider the $\IR$-bundle
$M:=(\IR\times\IR^3)\times\IR=\left\{(t,{\bf r},s)\right\}$  [2]. The
the Bargmann group acts on $M$ according to (1a-b), augmented by
$$
s^*=s-{\bf b.}A{\bf r}-\2 t{\bf b}^2-h,\qquad h\in\IR.
\eqno (1c)
$$

The action (2) on the wave functions can then be recovered by lifting
the wavefunctions to the bundle as equivariant functions i.e. by
replacing $\psi({\bf r},t)$ by $\Psi({\bf r}, t, s)
:=e^{ims/\hbar}\psi({\bf r}, t)$.

The idea of Duval et al. [3] is to view the extended
manifold $M$ as the proper arena for classical mechanics. Potentials of the
Newtonian type can be incorporated into a Lorentz metric defined on 
the extended space;
the dynamical trajectories  in ordinary space then correspond to {\it
null-geodesics} in the extended space.  This has been noted many years ago by
Eisenhart [4], but has long been  forgotten.
Extending to $M$ has the additional advantage the resulting
quantities become invariant [2].

The construction is reminiscent of Kaluza-Klein theory, except for that
the extra dimension
is null rather than space-like. Notice also that, while Kaluza-Klein
theory involves electromagnetism, the present framework is adapted to
(non-relativistic) gravitational interactions.

\chapter{Classical dynamics.}

Let us first consider a free, non-relativistic particle with Lagrangian
  $L_0=\2 m({\bf r}')^2$, ($(\ .\ )'={d\over dt}$). A Galilei transformation
  (1a-b)
  changes $L_0$ as
$
L_0\mapsto L_0+m({\bf b}^2/2+{\bf b.}A{\bf r}').
$
The non-invariance of the Lagrangian can however be compensated by
adding a fifth
coordinate, $s$, and by considering rather
$$
{\cal L}_0=L_0+m {ds\over dt},
\eqno (3)
$$
which is indeed invariant with respect to the action (1a-c) of the Bargmann
group.

Remarkably, the new Lagrangian ${\cal L}_0$ is
associated with {\it geodesic motion} in extended space,
$$
{\cal L}_0 =\2 m\left(({d{\bf r}\over dt}\big)^2+2{ds\over dt}\right)=
\2 mg^0_{ab}{x'}^a{x'}^b,
\qquad (a,b = 1,\ldots, 5)
\eqno (4)
$$
for the 5-metric
$
g^0_{ab}dx^adx^b=d{\bf r}^2+2dsdt.
$

Potentials can be
included at this stage. If $U({\bf r}, t)$ is a potential function, we
can consider the modified metric
$$
g_{ab}dx^adx^b=g^0_{ab}dx^adx^b-2Udt^2\equiv d{\bf r}^2 + 2dsdt-2Udt^2.
\eqno (5)
$$
Its extremals (geodesics in 5 dimensions) are conveniently described in
a homogeneous framework, i.e.
by the Lagrangian
$$
{\cal L}({\bf r},t,s,
\dot{{\bf r}},\dot{t},\dot{s})=
\left(
\2m g_{ab}{{\dot x}^a\over {\dot t}}{{\dot x}^b\over{\dot t}}\right)
{\dot t}\equiv
\2{m\over{\dot t}} g_{ij}{\dot r}^i{\dot r}^j
-mU{\dot t}
+m{\dot s},
\eqno (6)
$$
where the ${\bf r},t,s,\dot{\bf r},\dot{t},\dot{s}$ are coordinates on the
tangent bundle $TM$.

The value of the quadratic quantity
$h_{0}=g_{ab}\dot{x}^a\dot{x}^b$ is conserved
along any geodesic, and is interpreted as (minus) the
internal energy of the particle.
It is convenient to restrict our attention to those ``motions''
in external space for which $h_{0}$ vanishes [3, 4], i. e. to consider
{\it null geodesics}.
As it is readily verified, these latter  project
onto the extremals in ordinary space of the Lagrangian
$
L=L_0-mU.
$
(The vertical coordinate satisfies $s(t)=s_0-\int \!L dt$). We describe
classical system henceforth by the
Lagrangian (6), supplemented with the constraint of having vanishing internal
energy $h_{0}=0$.

The metric in Eq. (5) is a Lorentz metric on $M$ with signature
$(+,+,+,+,-)$, which admits a covariantly constant Killing vector,
namely
$
\xi=\partial/\partial s.
$
Following Ref. 3, $(M, g, \xi)$ is called a {\it Bargmann manifold}.

At the quantum level, the  Schr\"odinger equation
$$
[-{\hbar}^2{\Delta\over2m}+mU]\psi=i\hbar\partial_t\psi
$$
where $\Delta$ is the
Laplacian on ordinary 3-space can be written as
$$
\Delta_g\Psi = 0,
\eqno (7)
$$
$\Delta_g$ being the Laplacian on $(M, g)$, and $\Psi=e^{(ims/\hbar)}\psi$.

\goodbreak
\chapter{Symmetries.}

The five-dimensional framework is particularly convenient to describe the
symmetries of the problem. Firstly, massless geodesics are
permuted by conformal transformations. We
should, however, insist on that the mass be conserved and hence we only
consider transformations which preserve the vertical vector $\xi$.
Thus, we consider those vectorfields $Y$ on $M$ which
satisfy
$$
\L_Yg=\lambda g,
\qquad
[Y,\xi]=0.
\eqno (8)
$$
(Killing vectors correspond to $\lambda =
0$). Transformations as in (8) preserve the geodesic Lagrangian, 
$\L_{\widetilde
Y}{\cal L}=\lambda {\cal L}$, where ${\widetilde Y}$ is the canonical 
lift of $Y$ to the tangent
bundle $TM$. The associated Noether quantities,
$$
C={\partial{\cal L}\over{\partial{\dot x^a}}}Y^a,
\eqno (9)
$$
are constants of the motion. In conventional terms,
a vectorfield $Y$ as in Eq. (8) projects onto a
vectorfield, $X$, on ordinary spacetime denoted by $Q\times\IR$, and also to
$\IR$,
the time axis.
Let
${\widetilde X}$ denote the canonical lift of $X$ to
$TQ\times\IR$.  The condition $\L_{\widetilde Y}{\cal L}=0$ means then that
$\L_{\widetilde X} L=m{d\over dt}K$
for some real function $K$. Thus, the usual Lagrangian $L$
changes by a total derivative, which is the definition of a symmetry.
In fact, $Y=(X,-K)$. For (6), the Noether quantity (9) reduces to
$$
C=\big({\partial L\over{\partial{x'}^i}}\big)X^i
-\big({\partial L\over{\partial {x'}^a}}{x'}^a-L\big)X^t-mK,
\eqno (10)
$$
which is the standard conserved quantity associated to the symmetry $X$.

The Killing vector $\xi$ is always
a symmetry for a Bargmann system; the
associated Noether quantity (10) is, by (6), just $m$, the mass.

For a free particle, for example, Eq. (8) yields a 13-dimensional algebra,
whose action $Y=(X^i,X^t,Y^s)$ on extended spacetime is
$$
\Big(\omegab\times{\bf r}+(\2\delta+\kappa t){\bf r} +
\betab t +\gamma,\,\kappa t^2 +\delta t +\epsilon,\,
-\big(\2\kappa\ r^2+\betab {\bf .r}+\eta)
\Big).
\eqno (11)
$$
Here the $11$ parameters $\omega\in so(3),\,\betab,{\bf
\gamma}\in\IR^3,\,\epsilon,\eta\in\IR$ generate the isometries,
  with $\omega$
representing rotations, ${\bf \beta}$ Galilei boosts, ${\bf\gamma}$
space-translations, $\epsilon$ time-translations, and
$\eta$ translations in the vertical direction They are
are readily recognized as the generators of the Bargmann group.
The two additional parameters
$\delta,\kappa\in\IR$ generate the dilatations and the expansions. Eq. (11)
is the (extended) Schr\"odinger algebra [5]. Eq. (10) yields the
associated conserved
quantities,
$$
\matrix{
{\bf L}&=&{\bf r}\times{\bf p}\hfill&\hbox{angular momentum}\hfill\ccr
{\bf g}&=&m({\bf r}-{\bf v}t)\hfill&\hbox{center of mass}\hfill\ccr
{\bf p}&=&m{\bf r}'\hfill&\hbox{momentum}\hfill\ccr
-E&=&-{\displaystyle{{\bf p}^2\over 2m}}\hfill&\hbox{energy}\hfill\ccr
m&\hfill&&\hbox{mass}\hfill\ccr
D&=&\2{\bf p.r}-tE\hfill&\hbox{dilatation}\hfill\ccr
K&=&t^2E+2tD-\2m{\bf r}^2\hfill&\hbox{expansion}\hfill\ccr
}\eqno (12)
$$
as expected.

Let us notice finally, that, due to the conformal invariance of the Laplacian,
the classical symmetries (9) are symmetries of the Schr\"odinger equation also.
(Historically, the Schr\"odinger group was discovered as the maximal invariance
group of the Schr\"odinger equation of a free, non-relativistic particle [5]).

Other examples and the extension to spin are described
in References [6] and [7].

I am indebted to C. Duval in
collaboration with whom many of these results were obtained.

\vskip6mm

\centerline{\bf References}
\reference
V. Bargmann, Ann. Math. {\bf 59}, 1 (1954).

\reference
D. Simms, {\it Projective representations, symplectic manifolds and
extensions of Lie algebras}, '69 Marseille Lectures, Preprint 69/P.300,
(unpublished); V. Aldaya and J. A. De Azcarraga, Phys. Lett. {\bf 121B}, 331
(1983); Int. J. Theor. Phys. {\bf 24}, 141 (1985); G. Marmo, G. Morandi, A.
Simoni and E. C. G. Sudarshan, Phys. Rev. {\bf D37}, 2196 (1988).

\reference
C. Duval, G. Burdet, H. P. K\"unzle and M. Perrin,
Phys. Rev. {\bf D31}, 1841 (1985); see also
W. M. Tulczyjew,
J. Geom. Phys. {\bf 2}, 93 (1985);
M. Omote, S. Kamefuchi, Y. Takahashi, Y. Ohnuki,
in {\it Symmetries in Science III}, Proc '88 Schloss Hofen Meeting,
Gruber and Iachello (eds), p. 323 Plenum : N. Y. (1989).

\reference
L. P. Eisenhart, Ann. of Math. (Ser 2) {\bf 30}, 541 (1929).

\reference
R.~Jackiw, Phys. Today {\bf 25}, 23 (1972);
U. Niederer, Helv. Phys. Acta {\bf 45}, 802 (1972);
C. R. Hagen, Phys. Rev. {\bf D5}, 377 (1972).

\reference
C. Duval, {\it Quelques proc\'edures g\'eometriques en dynamique des
particules}, Th\`ese de Doctorat-\`es-Sciences, Marseille (1982)
(unpublished);
in {\it Conformal structures}, Proc.'85 Clausthal Conf.,
Barut and Doebner (eds), Springer
LNP {\bf 261}, p. 162  (1986);
C. Duval, G. Gibbons and P. A. Horv\'athy,
Phys. Rev. {\bf D43}, 3907 (1991).

\reference
C. Duval, in {\it Proc. XIVth Int. Conf. Diff. Geom. Meths. in Math. Phys.},
Salamanca '85, Garcia, P\'erez-Rend\'on (eds). Springer LNM
{\bf 1251}, p. 205 Berlin (1987);
J. Gomis and Novell, Phys. Rev. {\bf D33}, 2212 (1986);
J. P. Gauntlett, J. Gomis and P. K. Townsend,
Phys. Lett. {\bf B248}, 288 (1990);
P. A. Horv\'athy, in {\it Proc. XXIth Int. Conf. on Diff. Geom. Meths. in
Phys.}, Nankai'92. Yang, Ge, Zhou
(eds), Int. Journ. Mod. Phys., Proc. Suppl. {\bf 3A} p. 339 (1993).

\vfill\eject
\bye